\pdfoutput=1

\documentclass[preprint,prb,showpacs,superscriptaddress]{revtex4-1}
\usepackage{amsmath}
\usepackage{graphicx}
\usepackage{amssymb}
\usepackage{times}
\usepackage{color}
\definecolor{lr}{rgb}{1.0,0.3,0.3}
\definecolor{dg}{rgb}{0.0,0.5,0.0}

\graphicspath{./}
 


\begin{document}

\title{High fidelity bi-directional nuclear qubit initialization in SiC}

\author{Viktor Iv\'ady}
\email{vikiv@ifm.liu.se}
\affiliation{Department of Physics, Chemistry and Biology, Link\"oping
  University, SE-581 83 Link\"oping, Sweden}
\affiliation{Wigner Research Centre for Physics, Hungarian Academy of Sciences,
  PO Box 49, H-1525, Budapest, Hungary}

\author{Paul V. Klimov}
\affiliation{Institute for Molecular Engineering, University of Chicago, Chicago, IL , USA}

\author{Kevin C. Miao}
\affiliation{Institute for Molecular Engineering, University of Chicago, Chicago, IL , USA}

\author{Abram L. Falk}
\affiliation{Institute for Molecular Engineering, University of Chicago, Chicago, IL , USA}
\affiliation{IBM T.\ J. Watson Research Center, Yorktown Heights, NY, USA}

\author{David J. Christle}
\affiliation{Institute for Molecular Engineering, University of Chicago, Chicago, IL , USA}

\author{Kriszti\'an Sz\'asz} 
\affiliation{Wigner Research Centre for Physics, Hungarian Academy of Sciences,
  PO Box 49, H-1525, Budapest, Hungary}

\author{Igor A. Abrikosov}
\affiliation{Department of Physics, Chemistry and Biology, Link\"oping
  University, SE-581 83 Link\"oping, Sweden}
\affiliation{Materials Modeling and Development Laboratory, National University of Science and Technology `MISIS', 119049 Moscow, Russia}
\affiliation{LACOMAS Laboratory, Tomsk State University, 634050 Tomsk, Russia}

\author{David D. Awschalom}
\affiliation{Institute for Molecular Engineering, University of Chicago, Chicago, IL , USA}

\author{Adam Gali} 
\email{gali.adam@wigner.mta.hu}
\affiliation{Wigner Research Centre for Physics, Hungarian Academy of Sciences,
  PO Box 49, H-1525, Budapest, Hungary}
\affiliation{Department of Atomic Physics, Budapest University of
  Technology and Economics, Budafoki \'ut 8., H-1111 Budapest,
  Hungary}

\date{\today}


\begin{abstract}
Dynamic nuclear polarization (DNP) is an attractive method for initializing nuclear spins that are strongly coupled to optically active electron spins because it functions at room temperature and does not require strong magnetic fields. In this Letter, we demonstrate that DNP, with near-unity polarization efficiency, can be generally realized in weakly coupled hybrid registers, and furthermore that the nuclear spin polarization can be completely reversed with only sub-Gauss magnetic field variations. This mechanism offers new avenues for DNP-based sensors and radio-frequency free control of nuclear qubits.
\end{abstract}
\maketitle



Solid state quantum information processing (QIP) is a rapidly developing field, with numerous attractive quantum bit (qubit) candidates. Some qubit candidates that have stood out in particular are those based on the electron spin of color-center defects \cite{Baranov2005,Gali11pss,Weber10,Koehl11,Soltamov12,NatMat14,Christle2014,Widmann2014}. Among these are the nitrogen-vacancy center (NV center) in diamond\cite{duPreez:1965} and the divacancy\cite{Koehl11} and silicon vacancy\cite{Baranov2005,Riedel2012} color centers in silicon carbide (SiC). They are attractive for QIP applications because they have long spin coherence times that persist up to room temperature\cite{Balasubramanian:NatMat2009,Christle2014,Falk2013,Koehl11} and because their spin can be optically initialized and read out\cite{Jelezko:PSSa2006,Awschalom:Nature2010,Robledo:Nature2011,Koehl11,Falk2013}. These electron spin qubits can couple to nuclear spin qubits to realize hybrid quantum registers that combine long nuclear spin coherence times with optical addressability. This approach has already been employed to demonstrate QIP protocols including quantum error correction\cite{ErrorCor}, quantum memory\cite{Dutt:Science2007,Cappellaro:PRL2009,Fuchs2011,Maurer2012}, and nuclear gyroscopes \cite{Ajoy2012, Ledbetter2012}. A key prerequisite for employing hybrid registers is that their electron and nuclear spin qubits must achieve a high initialization fidelity. Achieving this high fidelity is especially difficult for nuclei that are only weakly coupled to an electronic spin.

In hybrid registers based on the NV center or the divacancies, nuclear spin qubits can be initialized optically via dynamic nuclear polarization (DNP)\cite{Jacques2009,Smeltzer2009,Fischer2013, FalkPRL2015,HaiJing2013,FischerPRL2013}. So far, most studies have focused on the excited-state DNP process, which utilizes the hyperfine coupling of the electron and nuclear spin in the electron's optically excited state\cite{Jacques2009,Smeltzer2009,Fischer2013, FalkPRL2015}. This pathway has led to $\sim99 $\% nuclear spin polarization for strongly coupled hybrid systems in both diamond\cite{Jacques2009} and SiC \cite{FalkPRL2015}. Ground state DNP, which can offer additional functionality compared to excited state DNP\cite{HaiJing2013,FischerPRL2013,IvadyDNP2015} including the DNP of weakly coupled nuclei, however, has been less explored in the context of QIP applications.

In this Letter, by using divacancy-based hybrid registers in SiC, we show that efficient ground-state DNP can be realized and finely magnetically controlled for weakly coupled nuclear spins. Our theoretical calculations demonstrate that the polarization of weakly coupled nuclei can exhibit a reversal from near $100$\% to $-100$\% for magnetic field variations as small as $0.3$ Gauss. We found this behavior to be true for half of the 300 hybrid registers that we considered, implying the generality of the mechanism. For strongly coupled nuclei we both experimentally and theoretically demonstrated the presence of a polarization reversal from $100$\% to $-25$\%. Our results indicate an avenue for sophisticated radio-frequency-free initialization and control of nuclear spin qubits and novel high-sensitivity dc-magnetometry protocols.

\begin{figure}[h!]
\includegraphics[width=0.95\columnwidth]{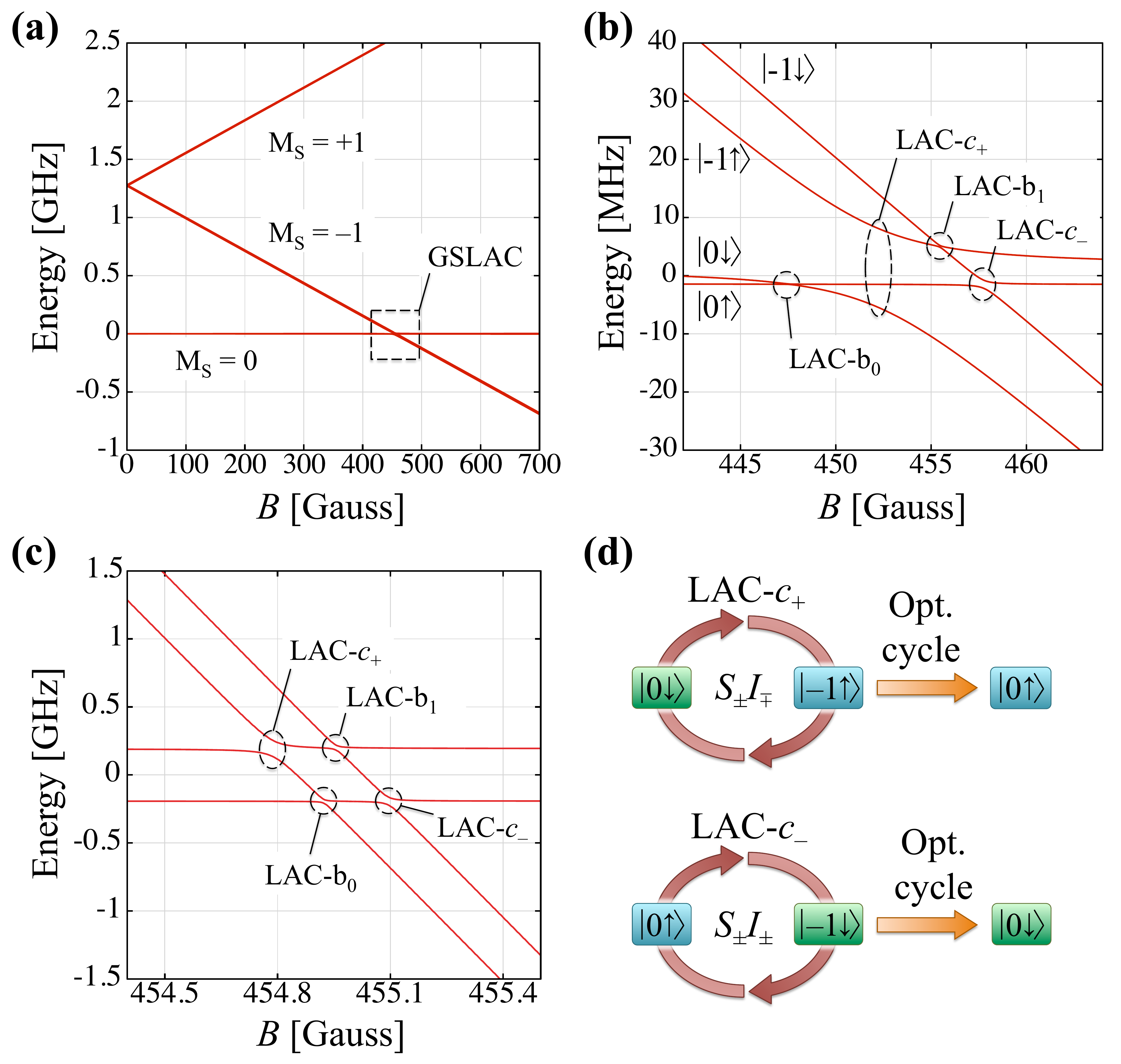}
\caption{ \label{fig:LAC} (Color online) Magnetic field dependence of the ground state fine and hyperfine structure of the 6H-SiC $hh$ divacancy and an adjacent $^{29}$Si nucleus. (a) Spin sublevels as a function of an axially applied magnetic field. Due to the zero-field splitting and the Zeeman effect, the $M_S = -1$ and $M_S = 0$ spin sublevels cross at $B_{\text{GSLAC}}$. (b) Magnified view of the GSLAC region for the case of $^{29}$Si nucleus at the Si$_{\text{IIb}}$ site\cite{FalkPRL2015,Son2006}. Due to the low symmetry of the hyperfine field at this site, four level-anticrossings form. (c) The case of a weakly coupled nucleus, when the nuclear Zeeman splitting determines the structure of the GSLAC region. $A_{\perp} \sim A_{\parallel} \sim 0.1$~MHz. (d) Schematic diagrams of the polarization processes at LAC-c$_{+}$ and LAC-c$_{-}$ that result in positive (upper chart) and negative (lower chart) nuclear spin polarization, respectively. }
\end{figure}

The neutral divacancy's spin can be polarized by optical excitation, owing to a spin selective non-radiative decay path from the $m_{s} = \pm 1$ manifold of the excited state to the $m_s = 0 $ spin sublevel of the ground state. In the DNP process of the divacancy \cite{FalkPRL2015,IvadyDNP2015}, the electron spin’s optical polarization cycle is linked to the nuclear spin states via hyperfine coupling, which allows repeated cycling to polarize nearby nuclear spins. At zero magnetic field, the large zero-field-splitting of the spin-$1$ state suppresses the weak hyperfine coupling of the electron and nuclear spins. On the other hand, by applying an appropriate magnetic field ($\pm B_{\text{LAC}}$) along the quantization axis of the divacancy, either the $m_s = +1$ or the $m_s = -1$ level becomes nearly degenerate with the $m_s = 0$ level, see Fig.~\ref{fig:LAC}(a), where the hyperfine interaction can effectively couple the electron and nuclear spins. Due to this interaction, small gaps open between the spin states and level anti-crossing (LAC) can be observed at $B_{\text{LAC}}$, see Fig.~\ref{fig:LAC}(b). As the divacancy's excited and ground states have similar fine structure, a LAC can be observed in both states. Due to the different zero field splitting in these states, $D_{\text{ES}}$ and $D_{\text{GS}}$, ESLAC and GSLAC occur at different magnetic fields, $\pm B_{\text{ESLAC}}$ and $\pm B_{\text{GSLAC}}$, respectively.

The fine and hyperfine structure of c-axis-oriented divacancy configurations' energy levels are described by the spin Hamiltonian
\begin{equation} \label{eq:GS_Ham}
\hat{H}_{\text{spin}} = D \left( \hat{S}_{z}^{2} - \frac{2}{3} \right)  + g_{\text{e}} \mu_{\text{B}} B_z \hat{S}_z  + g_{N} \mu_{N} B_z \hat{I}_z + \hat{\mathbf{S}}^{\text{T}} \mathbf{A} \hat{\mathbf{I}}\text{,}
\end{equation}
where  $\hat{\mathbf{S}}$ and $\hat{\mathbf{I}}$ are the electron and nuclear spin operators, $\hat{S}_z$ and $\hat{I}_z$ are the spin $z$ operators, $D$ is the zero-field-splitting parameter in the triplet state, $\mathbf{A}$ is the tensors of hyperfine interaction, $B_z$ is the external magnetic field parallel to the axis of the defect, $g_{\text{e}}$ and $g_{N}$ are the $g$-factor of the electron and nuclear spins, and $\mu_{\text{B}}$ and $\mu_{N}$ are the Bohr and nuclear magnetons, respectively. The spin Hamiltonian given in Eq.~\ref{eq:GS_Ham} can be applied for both the ground and excited states, however, $D$ and $\mathbf{A}$ are different in the two states.

In SiC, the most common paramagnetic nuclei are $^{29}$Si (4.8\% natural abundance) and $^{13}$C (1.1\% natural abundance), both of which have $I=1/2$ spin. Therefore, in the rest of this article we consider only spin-1/2 nuclei. The coupling of the electron and nuclear spins are described by the last term on the r.h.s.\ of Eq.~(\ref{eq:GS_Ham}). In general, the hyperfine tensor $\mathbf{A}$ can be parameterized by its eigenvalues, $A_{xx}$, $A_{yy}$, and $A_{zz}$, and the direction of the third eigenvector, which can be specified by the polar and azimuthal angles $\theta$ and $\phi$, respectively. In most cases $A_{xx} \approx A_{yy}$ and therefore the $\phi$ dependence can be neglected. Hereinafter, we use the following three parameters, $A_{\parallel} = A_{zz}$, $A_{\perp} = A_{xx} \approx A_{yy}$, and $\theta$. For positive values of the magnetic field, one can consider only the $\left| 0 \uparrow \right\rangle$, $\left| 0 \downarrow \right\rangle$, $\left| -1 \uparrow \right\rangle$, and $\left| -1 \downarrow \right\rangle$ states, in which basis the hyperfine Hamiltonian term in Eq.~(\ref{eq:GS_Ham}) can be written as\cite{IvadyDNP2015}
\begin{equation} \label{eq:hypop}
\hat{H}_{\text{hyp}} = \hat{\mathbf{S}}^{\text{T}} \mathbf{A} \hat{\mathbf{I}} = \frac{1}{2}
\begin{pmatrix}
 0 & 0 & \frac{1}{\sqrt{2}} b & \frac{1}{\sqrt{2}} c_{-} \\
 0 & 0 & \frac{1}{\sqrt{2}}  c_{+} & -\frac{1}{\sqrt{2}} b \\
\frac{1}{\sqrt{2}} b & \frac{1}{\sqrt{2}}  c_{+} & -A_z & -b \\
\frac{1}{\sqrt{2}} c_{-} & -\frac{1}{\sqrt{2}} b & -b & A_z \\
 \end{pmatrix}
 \text{,}
\end{equation}
 where
 \begin{eqnarray}
 A_z =  A_{\parallel} \cos^2 \theta  + A_{\perp} \sin^2 \theta  \\
 b = \left( A_{\parallel} - A_{\perp} \right) \cos \theta \sin \theta  \\
 c_{\pm} =  A_{\parallel} \sin^2 \theta   + A_{\perp}  \left( \cos^2 \theta  \pm 1  \right) \text{.}
 \end{eqnarray}

When the nucleus is located on the symmetry axis of the defect, e.g.\ the nitrogen of the NV center, the hyperfine field is symmetric, i.e.\ $\theta = 0$. The only non-zero off-diagonal element is $c_{+}$, which corresponds to the $\hat{S}_{\pm} \hat{I}_{\mp}$ operator combinations. In this case, only  $\left| 0 \downarrow \right\rangle \leftrightarrow \left| -1 \uparrow \right\rangle$ spin transition can be observed at LAC\cite{Jacques2009,Smeltzer2009}. On the other hand, in the general case, when the nucleus is not located on the axis of the defect, the symmetry of the hyperfine field is reduced ( $\theta \neq 0$). Therefore, other off-diagonal hyperfine coupling terms appear. These elements introduce other spin flipping processes that cause LACs at the crossings of other spin sublevels. In Fig.~\ref{fig:LAC}(b), four distinct LACs can be observed, which we label by LAC-$b_{0}$, LAC-$c_{+}$, LAC-$b_{1}$, and LAC-$c_{-}$, after the matrix elements that are responsible for the LACs. We note that LAC-$b_{0}$ occurs only if the $g$-factor $g_{N}$ of the nucleus is negative, for instance for $^{29}$Si, when the $\left| 0 \downarrow \right\rangle$ state is higher in energy than the $\left| 0 \uparrow \right\rangle$ state, thus they can cross each other, see Fig.~\ref{fig:LAC}(b). Furthermore, LAC-$b_{1}$ occurs only if $\left( A_z/2 - g_{N} \mu_{N} B_z \right)$ is positive, thus $\left| -1 \downarrow \right\rangle$ is higher in energy than $\left| -1 \uparrow \right\rangle$ and they cross each other too. In these cases, anticrossings occur due to non-zero $b$ related off-diagonal elements in the Hamiltonian matrix, see Eq.~\ref{eq:hypop}, that correspond to the precession of the electron and nuclear spins. On the other hand, in the general case, LAC-$c_{+}$ and LAC-$c_{-}$ take place and these LACs are connected to the off-diagonal matrix elements $c_{+}$ and $c_{-}$, which causes positive and negative nuclear spin polarization, respectively, via the processes depicted schematically in Fig.~\ref{fig:LAC}(d).

The different LACs correspond to different spin flip-flop processes, each of which is resonantly enhanced when the magnetic field passes through them. Therefore, the steady-state DNP can exhibit a complicated magnetic field dependence. In particular, the always present LAC-$c_{-}$ and LAC-$c_{+}$ processes cause positive and negative nuclear spin polarization at different magnetic fields. These fields, $B_{\text{LAC-}c_{\pm}}$, can be determined from the intersection of the energy levels as
\begin{equation}\label{eq:lacpos}
B_{\text{LAC-}c_{\pm}} = \frac{D \mp \frac{A_z}{2}} { g_{\text{e}} \mu_{\text{B}} \mp g_{N} \mu_{N}} \text{.}
\end{equation}

Note that, for sufficiently small values of the hyperfine splitting $A_z$ ($A_z < g_{N} \mu_{N} B_{\text{LAC}} \approx 0.4$~MHz), the nuclear Zeeman effect is the dominant interaction at the GSLAC. It determines the position of the energy level crossings, and thus the positions $B_{\text{LAC-}c_{\pm}}$ of LAC-$c_{\pm}$, see Fig.~\ref{fig:LAC}(c) and Eq.~\ref{eq:lacpos}. Importantly, for different weakly coupled nuclei of the same nuclear $g$-factor, these LACs occur at nearly the same magnetic fields, due the similar magnetic field splittings. On the other hand, as the nuclear Zeeman splitting strongly depends on the nuclear $g$-factor, the positions of the LACs are also determined by this factor. Here, we note that the $g$-factor of $^{13}$C and $^{29}$Si have different sign, therefore the positions $B_{\text{LAC-}c_{+}}$ and $B_{\text{LAC-}c_{-}}$ of LAC-$c_{+}$ and LAC-$c_{-}$ for the two nuclei tend to interchange.

For a given external condition, determining the nuclear spin-lattice relaxation time $T_1$, the steady state polarization largely depends on the overlap of the different LACs, which is equivalent to the overlap of different spin flip-flop processes. For the case of weakly coupled nuclei, the LACs are narrow, due to the weak hyperfine interaction. However, they are well separated by the nuclear Zeeman splitting in all cases. Consequently, the overlap of the two always present LACs, LAC-$c_{\pm}$, is greatly reduced, see Fig.~\ref{fig:LAC}(c). Therefore, based on the above considerations, we predict that the DNP of weakly coupled nuclear spins can give rise to efficient positive and negative polarization at well-defined magnetic fields. 

Furthermore, it was recently shown that in ESLAC DNP, the short optical lifetime and coherence time of the excited state (a few ns) suppresses slow spin flip-flop processes. ESLAC DNP is thus dominated by the fastest nuclear spin rotation process\cite{IvadyDNP2015}. The overall decay time is much longer for GSLAC DNP, which means that slower spin flip-flop processes can have a significant effect, e.g.\ in the case of a weak hyperfine interaction. Consequently, efficient DNP of weakly coupled nuclei is possible only in the GSLAC region.

To justify our hypothesis, we use a recently developed DNP model\cite{IvadyDNP2015}, as parameterized by \emph{ab initio} supercell hyperfine tensor calculations, to simulate the DNP of numerous $^{29}$Si and $^{13}$C nuclei around the $hh$ divacancy in 6H-SiC. In the calculations we considered all those sites for which $\left| A_{\perp} \right| + \left| A_{\parallel} \right|  > 100$~kHz. This criterion defines a sphere of $\sim\!\!10$~\AA\ that fits into our \emph{ab initio} simulations box. Sites of smaller hyperfine couplings are not considered due to size constraints and numerical accuracy issues of the \emph{ab initio} calculations\footnote{See Supplemental Material at [URL]}. 

\begin{figure}[h!]
\includegraphics[width=0.75\columnwidth]{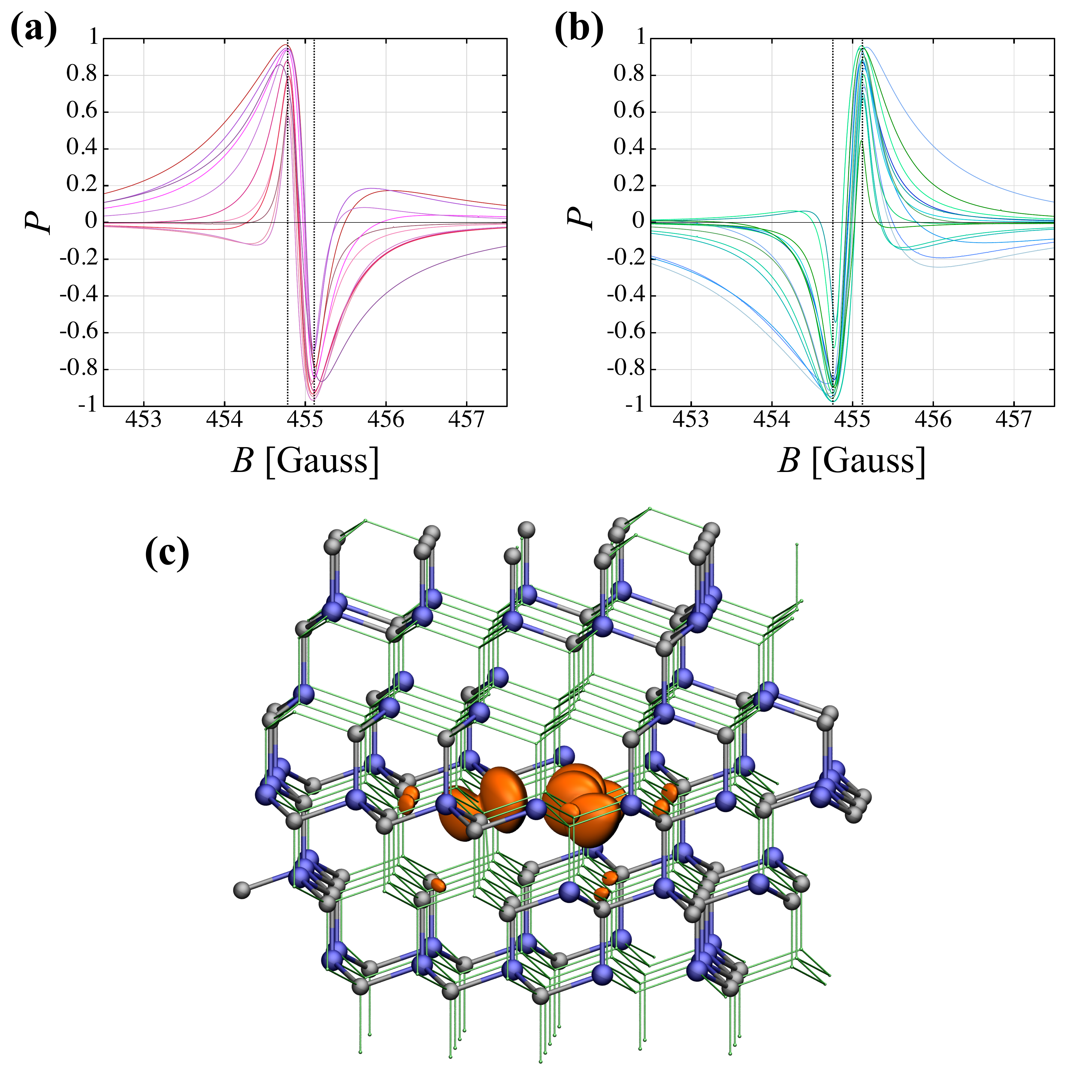}
\caption{ \label{fig:pol-dep} Magnetic field dependence of the DNP of (a) $^{29}$Si and (b) $^{13}$C nuclei at different  (c) weakly coupled neighboring sites around the $hh$ divacancy in 6H-SiC. Most of the polarization curves on (a) and (b) show ``peak and dip'' structure with a strong magnetic field dependence. Due to the opposite sign of the nuclear $g$-factor of $^{29}$Si and $^{13}$C, the order of the peak and the dip is opposite for the two nuclei (see text). (c) Weakly coupled $^{13}$C and $^{29}$Si sites around a $hh$ divacancy in 6H-SiC which exhibit well-developed peak and dip in their polarization curve. The orange lobes show the spin density of the divacancy that localized on the silicon vacancy site. The greenish wire frame shows the domain around the divacancy, in which the DNP calculations are carried out. The radius of the domain is maximized by the size of the simulation box used in the first principles calculations to obtain the hyperfine coupling tensors. Those Si and C sites that show "peak and dip'' polarization curve are represented by gray and blue balls, respectively. These sites are situated on an approximately spherical shell around the silicon vacancy site.}
\end{figure}

The simulated polarization curves are depicted in Fig.~\ref{fig:pol-dep}(a)-(b). As can be seen, the nuclear spin at numerous $^{29}$Si and $^{13}$C sites around a divacancy can be polarized with near 100\% certainty in both the $\left|\uparrow \right\rangle$ and $\left|\downarrow \right\rangle$ spin states. The positions of the polarization peak and dip are well defined according to Eq.~\ref{eq:lacpos}, and separated only by $\sim0.3$~Gauss. Among the considered 300 proximate sites, we find 11 and 16 symmetrically non-equivalent $^{29}$Si and $^{13}$C sites, respectively, for which exhibit a developed ``peak and dip'' like polarization curve. The corresponding 144 sites (60 $^{29}$Si and 84 $^{13}$C) are depicted in Fig.~\ref{fig:pol-dep}(c), which are located on an approximately spherical shell around the silicon vacancy site of the $hh$ divacancy. We emphasize that nearly 50\% of the considered nuclei can be initiated both in the $\left| \uparrow \right\rangle$ and $\left| \downarrow \right\rangle$ state with near unity fidelity by GSLAC DNP process, which demonstrates the generality of this mechanism. For the rest of the nuclei (not shown), either the hyperfine interaction was found to be too strong, resulting in the overlap of the LACs, or the $c_{-}$ off-diagonal element was found to be too small, and thus no polarization dip emerges. 

To experimentally investigate the existence of polarization inversion in the GSLAC DNP of divacancies, we carry out high resolution nuclear spin polarization measurements at the GSLAC region on an ensemble of $^{29}$Si nuclei in the Si$_{\text{IIb}}$ site\cite{FalkPRL2015,Son2006} of PL6 room temperature qubit in 4H-SiC. The $c$-axis-oriented PL6 color center's microscopic structure is still unknown, however, its similar spin and optical properties indicate close relationship to the neutral divacancies\cite{Koehl11,Falk2013,Falk2014}.  Furthermore, we use our DNP model to support and understand the observations.\cite{Note1}

\begin{figure}[h!]
\includegraphics[width=0.85\columnwidth]{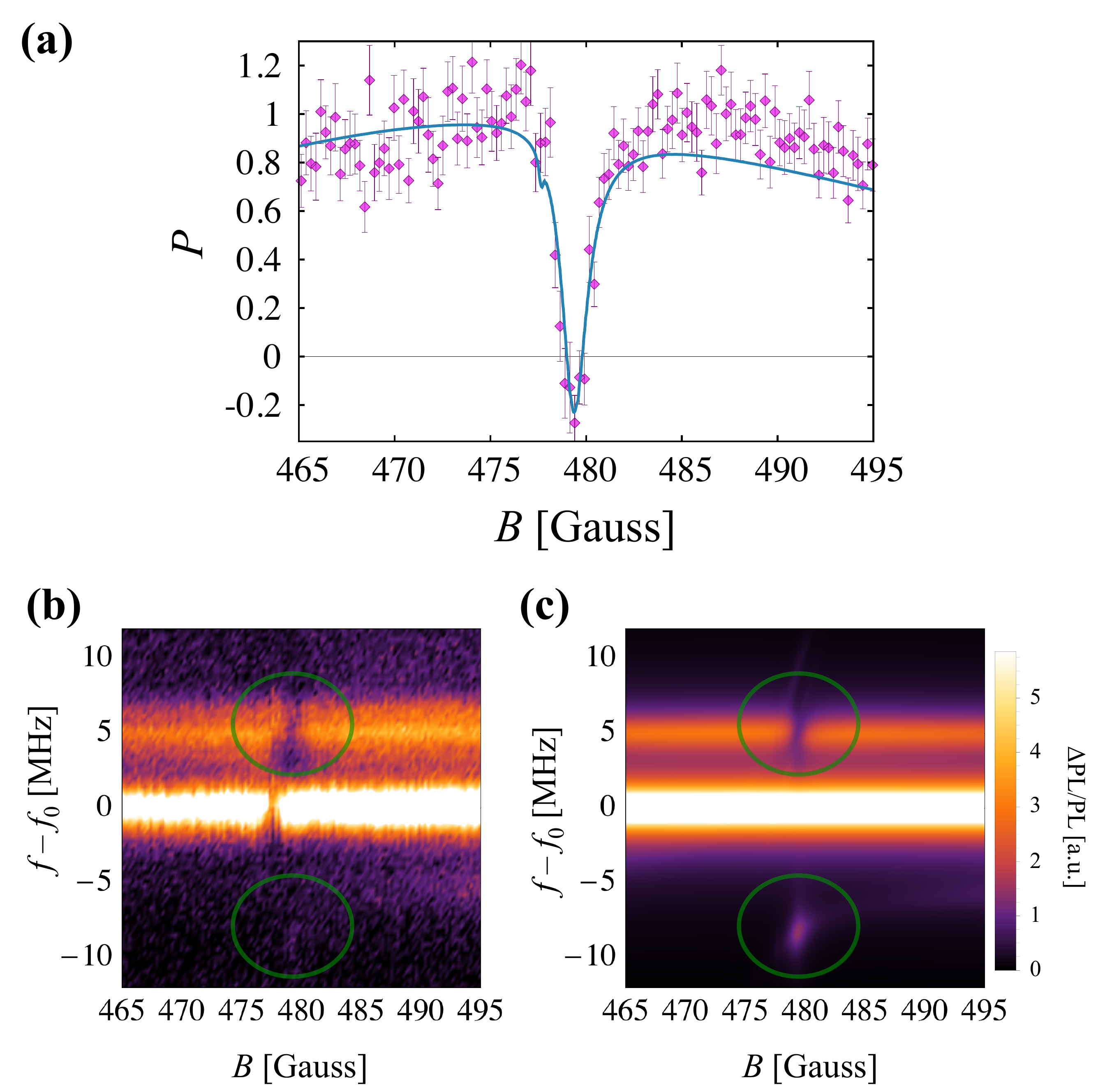}
\caption{ \label{fig:fit-PL6} Experimental and theoretical $^{29}$Si nuclear spin polarization and ODMR spectrum of the $M_S=0$ to $M_S=+1$ spin transition of PL6 qubits in 4H-SiC at the GSLAC region. (a) The measured (points) and calculated (thick line) magnetic field dependence of the nuclear spin polarization of a $^{29}$Si nucleus at the Si$_{\text{IIb}}$ site. DNP is highly efficient up until the LAC-$c_{-}$, at which point it exhibits a sharp drop and reversal.  Measurements are carried out at room temperature. For further details of the measurements and theoretical simulation see the Supplementary\cite{Note1}. (b) The experimental low-microwave-power ODMR spectrum. The measurements are carried out on an ensemble of PL6 divacancy relate qubits in 4H-SiC at room temperature. $f_0 = f_0 \! \left( B \right)$ describes the zero-filed-splitting and Zeeman shift of the $M_S=+1$ spin state. (c) Theoretical simulation of the ODMR spectrum, takes into account the DNP of the $^{29}$Si nucleus at the Si$_{\text{IIb}}$ site and the microwave transition strength in the M$_S=\left\lbrace 0, +1\right\rbrace$ manifold. The green ellipsoids on (b) and (c) highlight the signs of the nuclear spin polarization reversal.}
\end{figure}

The results of the measurements and theoretical calculations are presented in Fig.~\ref{fig:fit-PL6}. The experiment and theory are in close agreement, showing a sharp polarization drop and reversal of strongly coupled $^{29}$Si nuclei, from near $100$\% to $ -25$\%, in the vicinity of the GSLAC. This nuclear polarization feature shows the most sensitive magnetic field dependence as well as the largest polarization drop reported so far \cite{HaiJing2013}. Note that, in the case of Si$_{\text{IIb}}$ sites, the $^{29}$Si nuclear spin is strongly coupled, thus the hyperfine interaction dominates the GSLAC region and causes overlapping LACs, see Fig.~\ref{fig:LAC}(b). The overlap of different spin rotation processes reduces the efficiency of the GSLAC DNP, particularly the negative polarization\cite{Note1}. Nevertheless, the excellent agreement between theory and experiment on $^{29}$Si$_{\text{IIb}}$ nuclear spin polarization around GSLAC supports our theory and its implications on weakly coupled nuclear spins.

The theoretical discussions presented above are not restricted to the case of the divacancy in SiC. They can be readily generalized to the NV center in diamond\cite{Note1}, and to other optically polarizable high spin ground state color centers and adjacent nuclear spins. Furthermore, the demonstrated phenomenon exhibits great potential in various applications. For instance, the steep decrease and increase of the nuclear spin polarization with respect to the variation of the external magnetic field may give rise to DNP-based, dc-magnetometry protocols\cite{HaiJing2013}. Furthermore, as can be seen in Fig.~\ref{fig:pol-dep}, to invert the DNP process of weakly coupled nuclei, a small variation of the magnetic field, $\sim 0.3$~Gauss, is sufficient. These magnetic field variations are small enough that they can be induced by the magnetization of proximate electron spins. For instance, the variation of the spin state of an additional divacancy spin that is 5.1-6.5~nm away would provide a sufficiently large magnetic field to invert the nuclear spin polarization\cite{Note1}. This type of electron-qubit-controlled nuclear-qubit initialization could be used in various QIP applications, e.g. radio frequency-free nuclear qubit initialization for quantum memories.

Finally, we point out a few technological requirements of such applications. The efficiency of the ground state DNP process sensitively depends on the misalignment of the magnetic field\cite{IvadyDNP2015}. Our calculations show that the polarization curves, depicted in Fig.~\ref{fig:pol-dep}, are completely destroyed for magnetic field misalignments of only 0.1$^{\circ}$ relative to the $c$ axis of SiC. The polarization curves recover if this misalignment is not larger than 0.02$^{\circ}$. Therefore, precise control of both the strength and the direction of the magnetic field is required. Additionally, the host crystal must be isotopically purified to reduce the number of nuclear spins coupled to the electron spin qubit in the GSLAC region. A nuclear spin concentration of 0.08-0.8\%, which has been achieved in SiC\cite{PurifiedSiC}, can provide sufficiently dilute nuclear spin bath, while still exhibiting high probability for finding nuclear spins with $A_{\perp} \sim 50$-$200$~kHz coupling strength\cite{Note1}. 

In summary, we predicted and demonstrated a general process that allows weakly coupled nuclei to be initialized with near unity fidelity into both the $\left| \uparrow \right\rangle$ and $\left| \downarrow \right\rangle$ states in the GSLAC region of important solid state qubits. We provided a detailed understanding of the underlying physics and showed that GSLAC DNP can be used for radio frequency-free, magnetic-field controlled, nuclear qubit initialization. We experimentally and theoretically demonstrated the existence of DNP reversals of $^{29}$Si nuclei coupled to divacancy's spin in SiC. These results suggest the incorporation of GSLAC DNP into future QIP and quantum sensing protocols.

\section*{Acknowledgments}

Support from the Knut \& Alice Wallenberg Foundation projects ``Isotopic Control for Ultimate Materials Properties'' and ``Strong Field Physics and New States of Matter'' 2014-2019 (COTXS), the Swedish Research Council (VR) Grant 2015-04391, the Swedish Foundation for Strategic Research program SRL grant No.\ 10-0026, the Swedish Government Strategic Research Area in Materials Science on Functional Materials at Link\"oping University (Faculty Grant SFO-Mat-LiU No 2009 00971), the Grant of the Ministry of Education and Science of the Russian Federation (grant No. 14.Y26.31.0005), the Tomsk State University Academic D. I. Mendeleev Fund Program (project No. 8.1.18.2015), the Swedish National Infrastructure for Computing Grants No.~SNIC 2013/1-331 and SNIC 2014/1-420, the ``Lend\"ulet program" of Hungarian Academy of Sciences, the AFOSR, AFOSR-MURI, ARO, NSF, and UChicago NSF-MRSEC is acknowledged.

\bibliographystyle{apsrev4-1}
\bibliography{references}

\end{document}